\documentclass[8.5pt,twoside,twocolumn]{article}
\usepackage[super,sort&compress,comma]{natbib}
\usepackage{mhchem}
\usepackage{times,mathptmx,color}
\usepackage{amssymb}
\usepackage{sectsty}
\usepackage{balance}

\usepackage{graphicx} 
\usepackage{lastpage}
\usepackage[format=plain,justification=raggedright,singlelinecheck=false,font=small,labelfont=bf,labelsep=space]{caption}
\usepackage{fancyhdr}
\usepackage{epstopdf}
\definecolor{darkgreen}{RGB}{0,139,0}
\definecolor{turqoise}{RGB}{64,224,208}
\definecolor{brown}{RGB}{210,105,30}

\definecolor{b}{rgb}{0,0,1.0}
\definecolor{r}{rgb}{1,0,0}
\definecolor{g}{rgb}{0,1,0}

\begin{document}

\thispagestyle{plain}
\fancypagestyle{plain}{
\fancyhead[L]{\includegraphics[height=8pt]{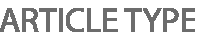}}
\fancyhead[C]{\hspace{-1cm}\includegraphics[height=20pt]{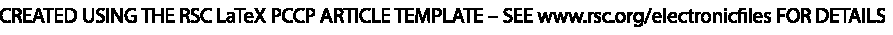}}
\fancyhead[R]{\includegraphics[height=10pt]{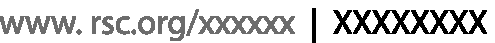}\vspace{-0.2cm}}
\renewcommand{\headrulewidth}{1pt}}
\renewcommand{\thefootnote}{\fnsymbol{footnote}}
\renewcommand\footnoterule{\vspace*{1pt}%
\hrule width 3.4in height 0.4pt \vspace*{5pt}}
\setcounter{secnumdepth}{5}

\makeatletter
\def\subsubsection{\@startsection{subsubsection}{3}{10pt}{-1.25ex plus -1ex minus -.1ex}{0ex plus 0ex}{\normalsize\bf}}
\def\paragraph{\@startsection{paragraph}{4}{10pt}{-1.25ex plus -1ex minus -.1ex}{0ex plus 0ex}{\normalsize\textit}}
\renewcommand\@biblabel[1]{#1}
\renewcommand\@makefntext[1]%
{\noindent\makebox[0pt][r]{\@thefnmark\,}#1}
\makeatother
\renewcommand{\figurename}{\small{Fig.}~}
\sectionfont{\large}
\subsectionfont{\normalsize}

\fancyfoot{}
\fancyfoot[LO,RE]{\vspace{-7pt}\includegraphics[height=9pt]{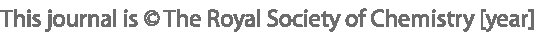}}
\fancyfoot[CO]{\vspace{-7.2pt}\hspace{12.2cm}\includegraphics{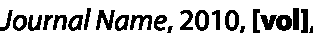}}
\fancyfoot[CE]{\vspace{-7.5pt}\hspace{-13.5cm}\includegraphics{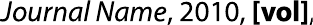}}
\fancyfoot[RO]{\footnotesize{\sffamily{1--\pageref{LastPage} ~\textbar  \hspace{2pt}\thepage}}}
\fancyfoot[LE]{\footnotesize{\sffamily{\thepage~\textbar\hspace{3.45cm} 1--\pageref{LastPage}}}}
\fancyhead{}
\renewcommand{\headrulewidth}{1pt}
\renewcommand{\footrulewidth}{1pt}
\setlength{\arrayrulewidth}{1pt}
\setlength{\columnsep}{6.5mm}
\setlength\bibsep{1pt}

\newcommand{\rev}[1]{{\color{red}#1}}

\twocolumn[
  \begin{@twocolumnfalse}
\noindent\LARGE{\textbf{Heaping, Secondary Flows and Broken Symmetry in Flows of Elongated Granular Particles}}
\vspace{0.6cm}

\noindent\large{\textbf{Geert Wortel,$^{\ast}$\textit{$^{a}$} Tam\'as B\"orzs\"onyi,\textit{$^{b}$} Ell\'ak Somfai,\textit{$^{b}$} Sandra Wegner,\textit{$^{c}$}, Bal\'azs Szab\'o,\textit{$^{b}$} Ralf Stannarius,\textit{$^{c}$} and Martin van Hecke\textit{$^{a}$}}}

\noindent\textit{\small{\textbf{Received Xth XXXXXXXXXX 20XX, Accepted Xth XXXXXXXXX 20XX\newline
First published on the web Xth XXXXXXXXXX 200X}}}

\noindent \textbf{\small{DOI: 10.1039/b000000x}}
\vspace{0.6cm}

\noindent \normalsize{In this paper we report experiments where we shear granular rods in split-bottom geometries, and find that a significant heap of height of least 40\% of the filling height can form at the particle surface. We show that heaping is caused by a significant secondary flow, absent for spherical particles. Flow reversal transiently reverses the secondary flow, leading to a quick collapse and slower regeneration of the heap. We present a symmetry argument and experimental data that show that the generation of the secondary flow is driven by a misalignment of the mean particle orientation with the {streamlines} of the flow. This general mechanism is expected to be important in all flows of sufficiently anisometric grains.
}
\vspace{0.5cm}
 \end{@twocolumnfalse}
  ]

\footnotetext{\textit{$^{a}$~Huygens-Kamerlingh Onnes Lab, Postbus 9504, 2300 RA Leiden, The Netherlands}}
\footnotetext{\textit{$^{b}$~Institute for Solid State Physics and Optics, Wigner Research Center for Physics,
Hungarian Academy of Sciences, P.O. Box 49, H-1525 Budapest, Hungary }}
\footnotetext{\textit{$^{c}$~Otto-von-Guericke-Universit\"at Magdeburg, Institute for Experimental Physics, D-39016 Magdeburg, Germany}}

\section{Motivation}
\subsection{Anisometric Granular Media}
Whereas most realistic granular media consist of non-spherical particles, many lab experiments and theories have focused on the behavior of near perfect granular spheres. Such a simplification is justified when the complex shape of the particles only has a limited, quantitative effect on the behavior. In contrast, here we unravel a general mechanism where non-spherical particles lead to strong secondary flows which cause significant heaping at the free surface of sheared granular media.

While the effect of anisometry has been studied very thoroughly for thermal particles such as in liquid crystals \cite{lc}, studies of the physics of anisometric,  athermal --- granular --- particles are more recent.
The packing density of random assemblies of aspherical particles
provides a striking
example of the subtle and significant effects of shape, as 
both prolate and oblate particles pack significantly more densely than spheres
\cite{philipse,torquato,review}.

What happens in flows? The packing density of flowing media is fundamental: 
for spherical granular media, the main coupling between micro structure and macroscopic mechanics, such as the resistance to flow, is through the local packing density \cite{reynolds}.  The situation for anisometric particles is potentially very rich: apart from causing either densification or dilation, flow can also lead to both ordering or disordering of the local packing, all of which in turn could influence the flow pattern. Open questions are thus:
What happens to the density of flowing anisomeric particles?
Does flow predominantly lead to ordering and densification, or does it mainly cause tumbling motion and concomitant strong dilation? How does ordering couple back to the flow? Evidence for both densification and dilation can be found in the literature: some experiments and simulations observe that elongated particle packings expand under shear~\cite{oda1,Herrmann, sandra14}, while others find that shear predominantly causes alignment and densification \cite{frette,campbell}. Clearly, the complex interplay of density, ordering, and flow is not well understood.

Here we report that the coupling between flow and ordering leads to the strong generation of a secondary flow. We perform experiments on the flow of rod-like particles in a split-bottom cell
without an inner cylinder \cite{fenistein1,fenistein2,dijksmansm,tamas,jaegersplibo}, a geometry in particular
well-suited for generating axial flow patterns at large filling heights \cite{fenistein2,tamas,jaegersplibo}.
We find the surprising formation of a considerable heap of grains at the free surface of the granular bed (see Fig.~1), in stark contrast to the flow of spherical grains where such a heap is completely absent.
We show that this heap formation is not a transient --- if the heap is removed, it rapidly reforms --- and argue that heaping is caused by secondary flows. For spherical grains, only very weak secondary flows have been observed \cite{jaegersplibo, saka,hill,wl}, but here the secondary flows are much stronger, as evidenced by surface observations and CT measurements.

What causes this flow, and what is the role of the {shape} anisotropy of the particles?
Reversal of the flow leads to a transient reversal of the secondary flow and
disappearance of the heap, after which it reforms. 
This suggests to consider the symmetries, and 
we provide evidence that misalignment between the mean particle orientation and the flow streamlines underlies the
secondary flow generation. 
We suggests that the mechanism observed here
could arise in a variety of granular flows of anisometric particles, and is thus a crucial ingredient in future theories and descriptions of realistic granular flows \cite{kamrin}.

\subsection{Heaping and Secondary Flows}

The phenomenon of secondary flow induced heaping in flowing anisometric grains presented here is novel. Nevertheless, both heaping and secondary flows have been observed in various granular media before, albeit in different combinations and situations than described here. To provide some perspective,
we below briefly outline what is known about the links between particle shape, heaping and secondary flows.

Heaping is well-known to arise when isometric granular media are {\em vibrated}. The heap formation is often driven by a weak vibration induced flow, typically caused by the difference between wall and internal friction \cite{heapvib}. Rod-shaped granular particles exhibit a wealth of additional phenomena when vibrated, where ordering of the particles in nematic-like states often plays a crucial role ~\cite{tamasrev}. For example, experiments in 3D on rods in a vibrated tube show that the rods align to the walls and form a high density nematic phase~\cite{villarruel,vandewalle,kudro2003,yonnel2013}; in
quasi-2D experiments similar alignment was found for dense packings ~\cite{nossal}.
At lower densities, the intricate coupling between local ordering, density and propulsion can lead to large fluctuations in the local density of active rod-like particles due to a competition between alignment and void formation ~\cite{narayan07}.

Secondary flows have been observed in several granular flows of isometric particles. For rapid granular flows in the 
liquid-like and collisional regimes, secondary flows are often generated through mechanisms akin to hydrodynamic instabilities
\cite{pouliquen,lin82,boyer1}. For dense, quasistatic flows as studied here,
weak secondary flows have also been observed for spherical beads, often related to the breaking of a symmetry. For example, recent experiments in Taylor-Couette flows showed that
reducing/increasing the gravity suppresses/enhances the convective like
secondary flow \cite{murdochPRL2013}, whereas
in hopper flows a slight perturbation of the axisymmetric geometry by using
a non-axisymmetric container or by
tilting it away from the vertical can also induce a secondary flow
\cite{gremaud2003}. 

Finally, flows of spherical particles in the split-bottom geometry as used here have shown the presence of a weak \emph{convection roll}, which can either lead to upwards or downwards motion in the core ~\cite{hill,wl,saka,jaegersplibo}.
These weak secondary flows do not create any significant amount of heaping (or formation of a dip). Recently we studied the flow of rods for {\em shallow filling heights} in the
split-bottom geometry and found that dilation dominates ordering in the flowing regions, and that the observed flow profiles are very similar to that of spherical grains --- no heaping was observed~\cite{sandra14}.

\begin{figure}[t]
\begin{center}
\includegraphics[width=\columnwidth]{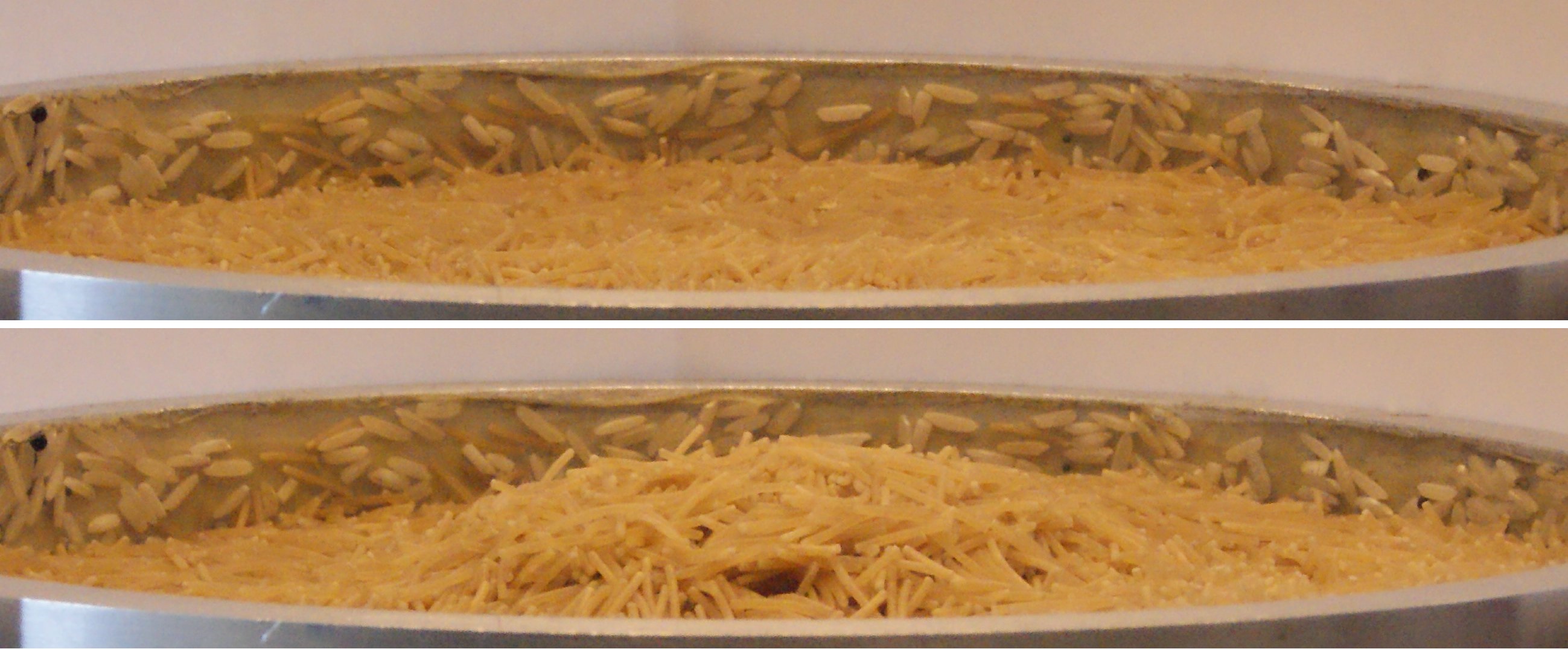}
\caption{{\footnotesize {Shear induced heaping of elongated "vermicelli" pasta grains in a cylindrical split-bottom
container.} Top: Initially flat surface. Bottom: Heap.}}\label{heap_pic}
\end{center}
\end{figure}

\section{Setup} \label{setup}
Our experiments are performed in a split-bottom flow geometry \cite{fenistein1,fenistein2,dijksmansm}. We note here that the ordering and secondary flow generation are likely not limited to split-bottom flows --- this geometry is merely experimentally convenient to obtain smooth, robust granular flows.
In this geometry, which is thoroughly studied for spherical particles, the granulate is poured in a cylindrical container, the bottom of which consists of an inner disk and outer ring. The relative motion of disk and ring then generate a wide shear band which emanates from the edge of the disk, away from the boundaries \cite{fenistein1,fenistein2,dijksmansm} --- being away from lateral boundaries is important for rod-like particles, as boundaries could cause ordering \cite{mueth}. The three-dimensional flow profile crucially depends on the relative filling height $h_0/R_s$, where $R_s$ is the radius of the inner bottom disk, and $h_0$ the filling height of the grains in the container.  For shallow layers ($h_0/R_s \lesssim 0.45$) the 3D shear zones form a trumpet-like shape, and the main shear direction is in the horizontal planes; for deeper layers, as is the focus here, the shear zone forms a continues dome, and axial shear becomes  important \cite{dijksmansm}.

We perform measurements in two different setups: $(i)$  a motorized metal split-bottom cell used to investigate the phenomenology of heaping by observing the free surface, and $(ii)$ a hand driven, plastic split-bottom is used in an X-ray CT scanner.

\begin{figure}[t]
\begin{center}
\includegraphics[width=6cm]{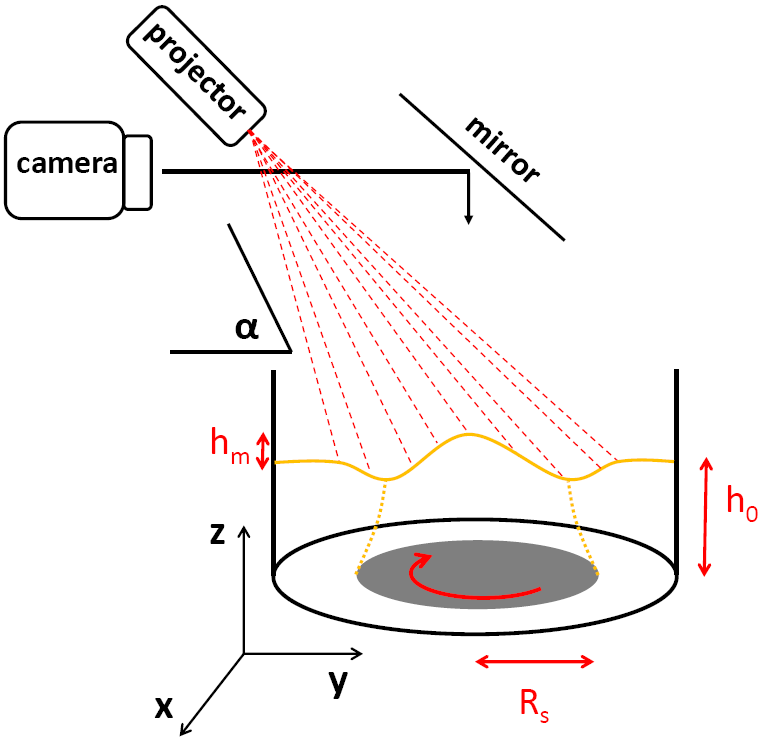}
\caption{{\footnotesize Schematic of the setup $(i)$, with radius of the {rotating} bottom disk $R_s$ and filling height $h_0$ indicated. Imaging of a series of lines that are projected onto the surface at an angle $\alpha$ allows to reconstruct the surface profile and
height of the heap. }}\label{setup1}
\end{center}
\end{figure}

{\em (i) Surface Measurements ---} Measurements of the heap formation at the free surface are carried out in a standard split-bottom cell with an outer radius of 110~mm and disk radius $R_s$ of 85~mm
(Fig. \ref{setup1}). The bottom disk is {connected} to duty-cycle controlled 24 V DC motor which drives the flow. This disk is visible from below, enabling us to measure the rotation rate. For all experiments, we use a rotation rate of 0.07~rps. The precise value of the rate is not crucial as the slow granular flow is rate independent. Rice grains are glued to the inside of the cell to ensure no-slip boundary conditions. We have performed experiments with a range of particles, and focus here
on ``Surinam rice'' grains, which have a long axis of approximately $7\pm1$~mm and short axes of 2.0 and 1.5~mm (aspect ratio $Q\approx4$), and ``vermicelli'' pasta grains with a diameter of 1~mm and length of 14 $\pm$ 3~mm (aspect ratio $Q\approx14$). These natural materials are more sensitive to wear and moisture than regular glass beads. Whereas measurements reproduce well over the cause of days, we did find larger fluctuations in measurements that were taken months apart:
even though the heaping phenomenon is robust,
the precise values of, e.g., maximum heap height have considerable scatter.

To measure the height profile of the free surface, we project a pattern of parallel lines on the surface using an Epson EB-824 projector, aimed at the surface under an angle $\alpha =$ 51$^\circ$; undulations of the lines observed from above correspond to variations of the surface height. To reconstruct the surface height, we record the pattern of distorted lines from above with a Foculus FO432B camera via a mirror. The spacing between the lines is 10~mm, and we divide the surface in a $10\times10$ mm$^2$ square grid, and obtain the local surface height $\tilde{h}$  via the deformation of the lines with a vertical accuracy of about $\pm 1$~mm.
In the remainder we focus on the change in surface height with respect to the filling height, and define $h(x,y,\theta)=\tilde{h}(x,y,\theta)-h_0$, where $h_0$ is the filling height $\langle h(x,y,\theta=0) \rangle$, {where $\theta$ is the rotation angle of the plate}.

{\em (ii) 3D Tomography Setup ---} The X-ray tomography experiments are performed in a split-bottom cell with an outer radius of 19.5~cm. Here the inner disk is attached just above the bottom of the container and has a radius of 13~cm and a thickness of 6~mm. In contrast to the surface measurements, the particles are sheared by rotating the outer wall --- for slow flows where inertia does not play a role, this leads to the same flowing regions as when rotating the inner disk \cite{dijksmansm,corwin}. The cell is filled to a filling height of $h_0/R_s$=0.54 with cylindrical wooden pegs with {length of $2.5$ cm and  diameter of $0.5$ cm} ($Q=5$). The scanner is a medical X-ray angiography machine (Siemens Artis zeego) at the INKA lab, Otto von Guericke University, Magdeburg. It consists of a rotational C-arm based X-ray source mounted on a high-precision robot-arm with a flat-panel detector featuring high resolution whole volume computer tomography scanning~\cite{sandra12}. We make a scan after each 1/16 of a rotation and obtain a resolution of 0.492 mm/voxel, which is accurate enough to identify individual particles.

\begin{figure}[t]
\begin{center}
\includegraphics[width=\columnwidth]{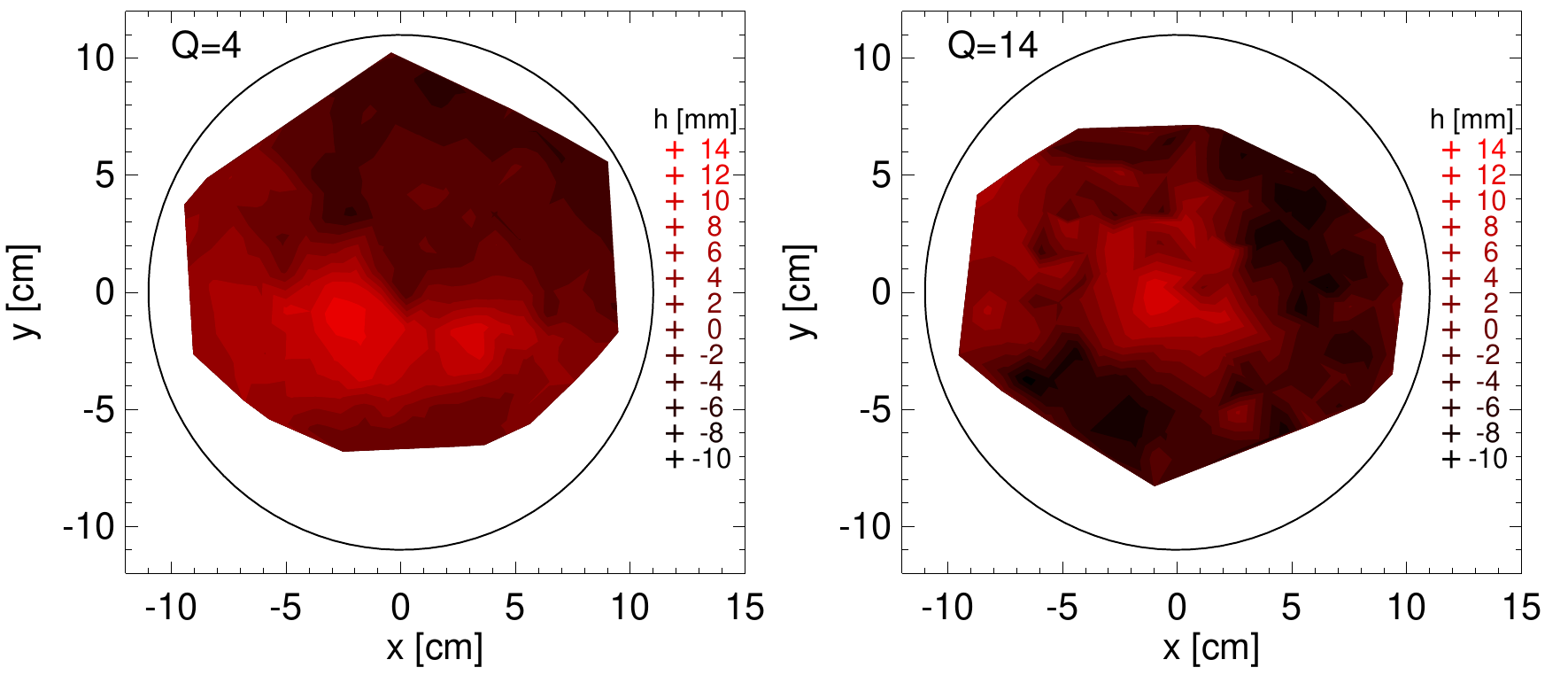}
\caption{{\footnotesize Vertical elevation of the surface after 100 revolutions for (a) $h_0/R_s=0.46$ and rice grains $Q=4$, and  (b) $h_0/R_s=0.55$ and pasta grains $Q=14$. The circle indicates the location of the sidewall --- peculiarities of the height imaging make that we cannot estimate elevations close to this edge.
}}\label{top}
\end{center}
\end{figure}

\section{Heap Formation} \label{r2d}

When we start shearing a packing of rods, we observe that the particles in the shear band align and that the packing globally expands. For shallow filling heights, the shear band remains localized above the edge of the disk, but for increasing
filling heights, the shear bands meet, the center of the packing rotates with
a different rate than the disk (we quantify this with the precession rate $\omega_p$ --- see \cite{fenistein2,tamas,jaegersplibo}), and a heap forms near the center of the cell. Measurements on various rod-like particles reveal that this behavior robustly occurs as long as the particles aspect ratio is $\geq 3$.

In Fig.~\ref{top} we show two-dimensional plots of the vertical elevation of the surface
in comparison to the initial height after l00 revolutions, for rice grains ($Q=4$) and pasta grains ($Q=14$). These plots illustrate that for $Q=4$ we typically find that the apex of the heap is off center, whereas for large $Q$ the apex is essentially in the center of the flow cell.
While its height quickly attains a fairly constant value, such heaps are clearly dynamic: we {observe} large scale fluctuations in their shape, and also clearly can observe particles avalanching down the slopes of the heap. This strongly suggests that a secondary flow drives the heaping process.

In the first set of quantitative experiments, we have studied the steady state height of this heap. We note here that there are several subtleties in these measurements. First, the height profile has a microscopic roughness comparable to the particles size, which is considerable. Second, in the deep flow regime the overall packing dilates considerably. Third, whereas in the initial state the rods typically align horizontally, during flow they tumble, and one can observe grains 'poking out of the surface', leading to an increase of the mean surface height, even in the absence of dilation.
We therefore define the heap height $h_m$ and deepest dip $h_l$ by the average over the five highest and lowest grid points $h(x,y)$, where heights are measured with respect to the initial filling height.
The highest points
are lying close together, near the top of the heap, whereas the lowest points {are typically} in a ring like-shape. In both cases
averaging minimizes measurement noise.

\begin{figure}[t]
\begin{center}
\includegraphics[width=\columnwidth]{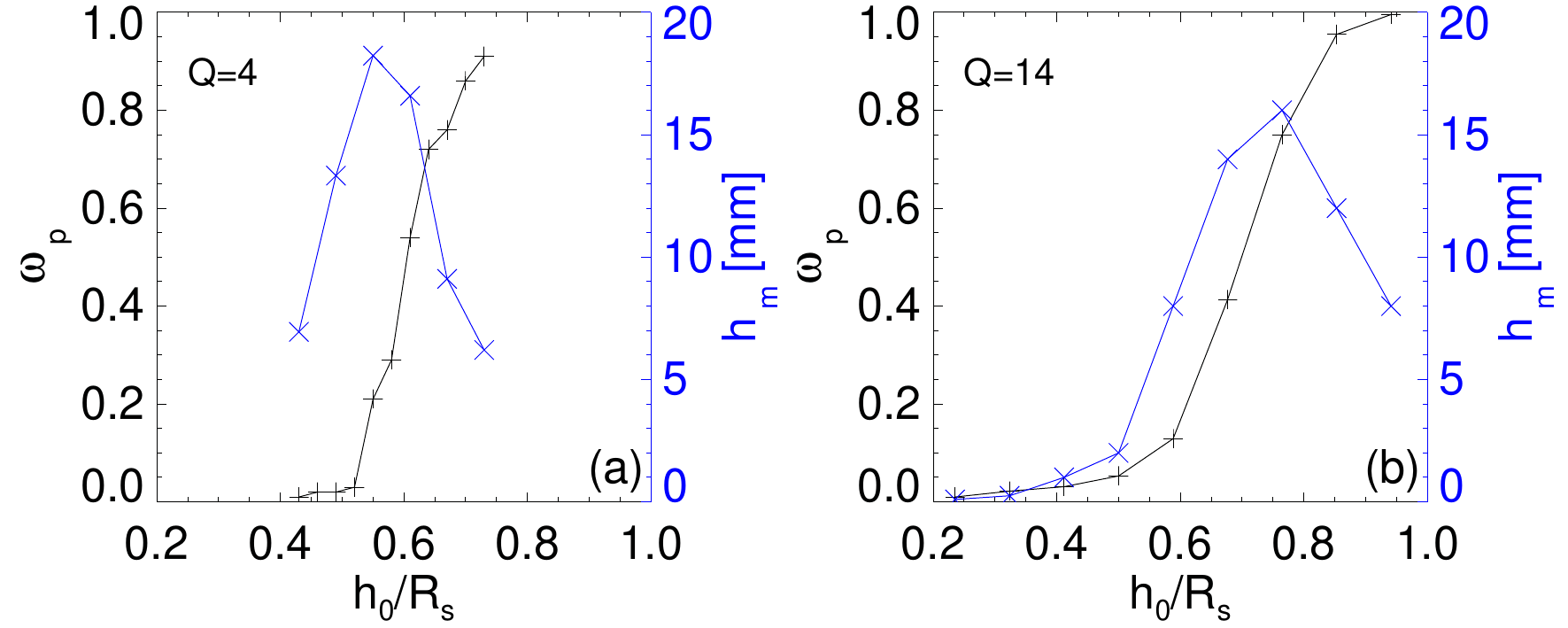}
\caption{{\footnotesize Heap height $\mathit{h_m}$ ($\times$) and precession $\mathit{\omega_p}$ (+) as a function of $h_0/R_s$ for rice grains $Q=4$ (a) and pasta grains $Q=14$ (b).}}\label{hdep}
\end{center}
\end{figure}

The filling height dependence, for two different aspect ratios, of the precession rate $\omega_p$
and $h_m$are shown in Fig.~\ref{hdep}. For both aspect ratios, we find that the precession rate monotonically increases with filling height --- in qualitative agreement with the precession of spherical particles \cite{fenistein2}. In contrast, the maximum heap height is non-monotonic with filling height. As expected, there is no significant heaping for low filling height. The heaping is strongest for intermediate filling heights, roughly in the regime where the precession is around 0.5; here the axial shear near the free surface is strongest. Note that for very tall filling heights where $\omega_p \rightarrow 1$,
the shear zone is deep below the free surface \cite{fenistein2,tamas,jaegersplibo,dijksmansm}, and no significant heaping is observed.

\begin{figure}[t]
\begin{center}
\includegraphics[width=7cm]{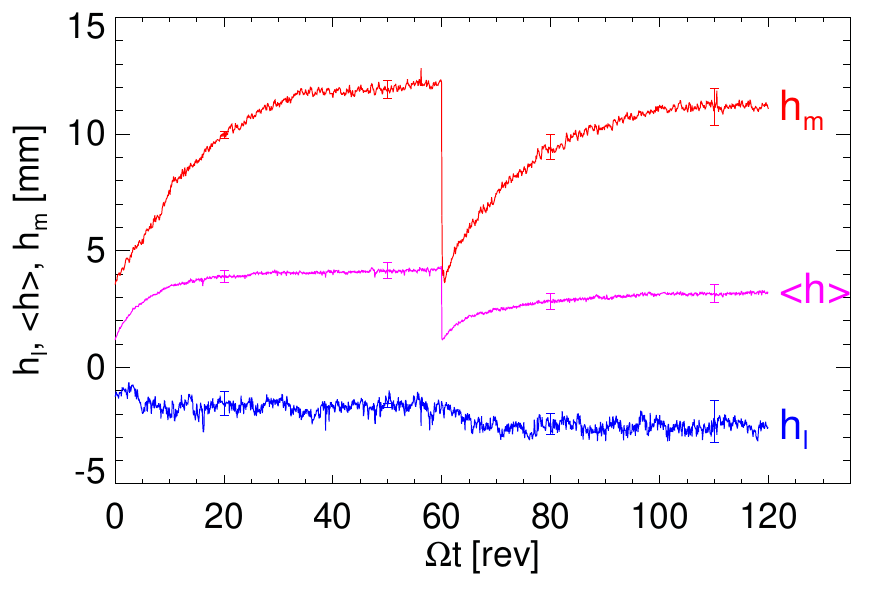}
\caption{{\footnotesize Growth and regrowth of a heap as function of the number of disk revolutions $\Omega t$ for rice grains $Q=4$, $h_0/R_s=0.49$. The data is an average over 10 independent runs, and at 60 {revolutions} the heap is removed (see text).
The heap height $h_m$ is plotted in red, purple is the average height $\langle h \rangle$ (we see that on average the system dilates) and blue is the lowest point $h_l$.
}}\label{removal}
\end{center}
\end{figure}

In principle, it would be possible that the heap only forms when starting out from a non-sheared, freshly poured granular assembly. To probe if and how the heap reforms in a presheared system, we performed experiments where we first shear the system for 60~rev  --- long enough to reach a steady heap state. We then stop the flow and remove the heap manually (with a suction device) without perturbing the rest of the packing.  We then restart the flow in the same direction and observe the reformation of the heap. In Fig.~\ref{removal} we show the evolution of the heap as function of the number of disk revolutions. The reported points are averaged over 10 runs where the standard deviation over the 10 {runs} is typically 0.5~mm. The main finding is that the formation rate and steady state height of the original and second heap are essentially equal, up to a small difference in height consistent with the removal of granular material at 60 {revolutions}. This
suggests that the driving mechanism for heap formation is
not sensitive to the difference between freshly poured and presheared systems.

\begin{figure}[t]
\begin{center}
\includegraphics[width=\columnwidth]{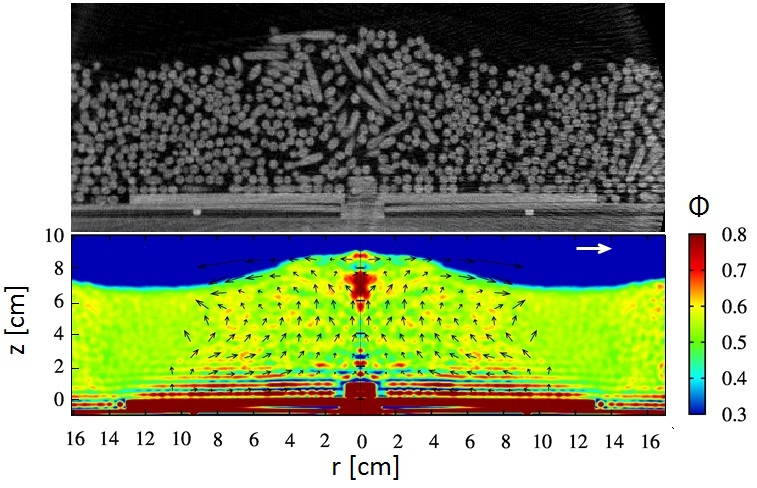}
\caption{{\footnotesize top: An example slice throught the center of the cell of the tomographic image. Individual particles wooden peg particles $(Q=5)$ can clearly be observed. bottom: Density $\Phi$ (color) and velocity (arrows) as a function of $r$ and $z$. The data is averaged over $\phi$ and 83 scans of the full system, which corresponds to approximately 5 {revolutions}, all in steady state. The density field shows that the density is slightly lower in the shear band than in the core. The velocity field shows a clear convective roll that moves the particles upwards for small $r$. For large $r$, we cannot {track particle motion because tangential} particle displacements in between two scans are too large to see which particle corresponds to which particle in between two frames. The white scale arrow corresponds to a convection speed of 0.055~cm/rev, which is $1.4\cdot10^{-3}$ times the {tangential} velocity of the grains - in the shear band just outside the inner disk - that corotate with the outer wall. }}\label{tomo_ex}
\end{center}
\end{figure}

\section{Secondary Flow} \label{r3d}
For granular flows in split bottom cells, a very weak secondary flow can be observed \cite{saka,hill,wl}. We will now show that a much stronger secondary flow, with significant {\em up flow} near the center, drives the growth of the heap.

To access the full 3D flow we perform experiments in an X-ray CT scanner.
We note that in this setup $(ii)$, the heaping effect arises in qualitatively the same manner as in setup $(i)$. In Fig.~\ref{tomo_ex}(a) we show an example image of a reconstructed vertical slice through the center of the measuring cell.  This image clearly shows that we can see each individual particle, allowing us to extract the local density, velocity field, and track the particles to obtain their precise orientation, and the {orientational order} tensor $T$~\cite{tamasPRL12,tamasPRE12,sandra12}.  Moreover, it is clear that the particle orientation exhibits spatial structure: in the shear bands above the disk edge, the main orientation is {tangential}, whereas near the middle of the cell, the particles mostly are vertical.

In Fig.~\ref{tomo_ex} (bottom panel) we show the density $\Phi$ (color) and velocity (arrows) as a function of $r$ and $z$. This data is averaged over $\mathit{\phi}$ and 83 scans of the full system, which corresponds to approximately 5 {revolutions}, all in steady state. The velocity field clearly shows convective rolls that transport the particles upwards at small $r$: heaping is the result of a secondary convective flow. The heap height is limited by the avalanching down of grains of the sides of the heap, as we observed at the free surface.
The density field reveals that the packing density is slightly lower in the shear band than in the core, suggesting that higher local strain rates favor dilation.

\begin{figure}[t]
\begin{center}
\includegraphics[width=7cm]{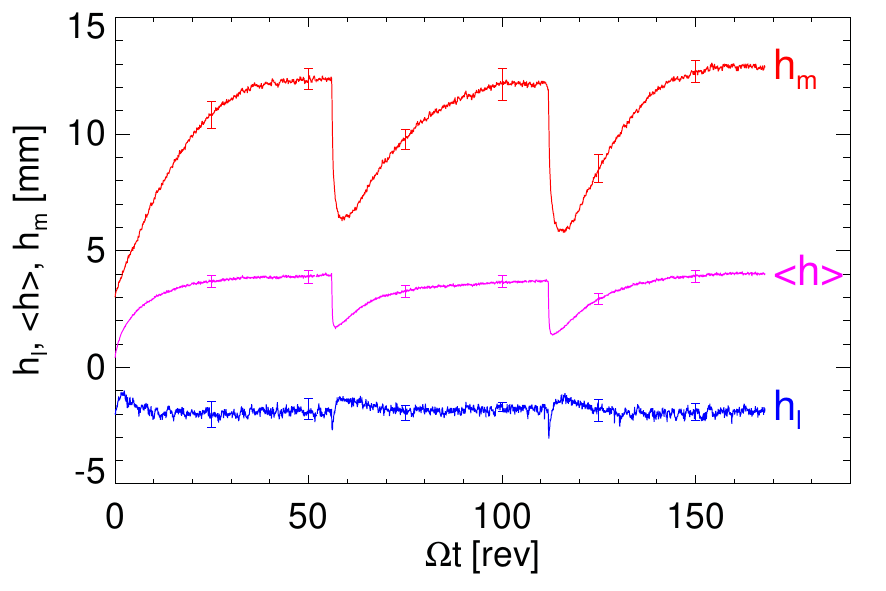}
\caption{{\footnotesize In this experiment we shear for 53 {revolutions}, then shear in the other direction for 53 {revolutions} and then reverse again and shear for 53 {revolutions}. Upon reversal, the heap first abruptly disappears and then grows back. This data is for rice grains, $Q=4$ and $h_0/R_s=0.49$.}}\label{reversal}
\end{center}
\end{figure}

What causes this secondary flow? Clearly, a coupling between the orientation and
ordering of the particles, primary flow, and secondary flow must play a role.
To get more insight in this coupling,
we have performed experiments where we, after reaching a steady state, reverse the flow direction. Surprisingly, after reversal, the heap quickly disappears and then reforms
at a rate comparable to growth from a freshly poured sample (Fig.~\ref{reversal}).  The same process is observed when we reverse the flow direction again. We conclude that upon reversal, the secondary flow direction reverses transiently --- consistent with this, we even observe a small dip at the center of the surface before the heap reforms.

The fact that the secondary flow can be temporarily reversed by reversing the primary flows direction, implies that
the direction of shear must be encoded in the fabric of the packing.
The absence of any significant secondary flow for spherical particles suggests that particle orientation is the dominant factor in setting the granular fabric here.

\begin{figure}[t]
\begin{center}
\includegraphics[width=\columnwidth]{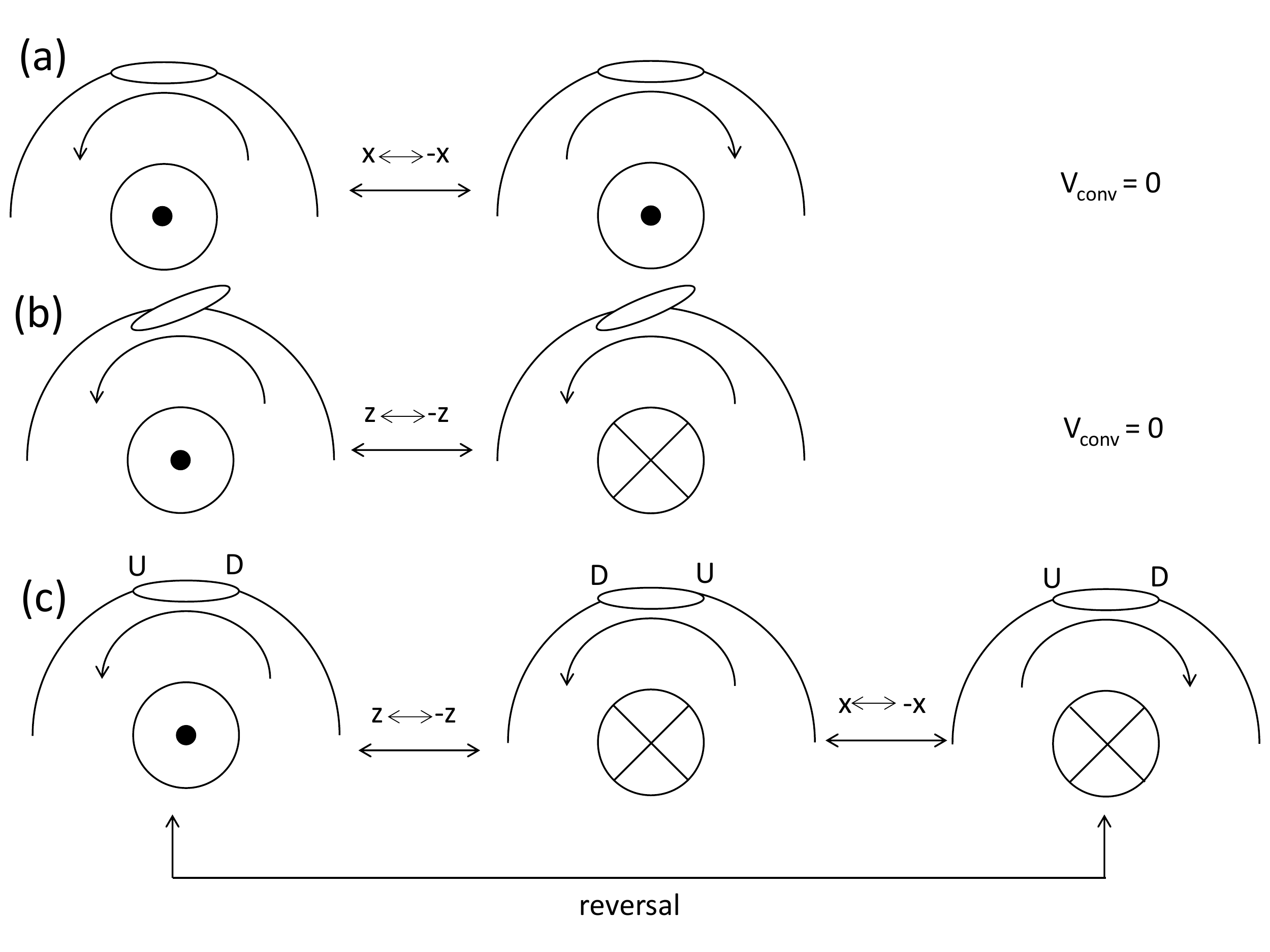}
\caption{{\footnotesize Using a schematic topview representation of the particle orientation (U: up and D:down), shear direction and convection direction (dot: up, crosses: down), we investigate which particle orientation effect is allowed to cause the heaping by symmetry (see text).}}\label{rice_refl}
\end{center}
\end{figure}

We will now consider the (approximate) symmetries of the system to unravel how particle orientation and secondary flow couple --- see Fig.~\ref{rice_refl}. We first consider the
reversal of the secondary flow, immediately after reversal of the primary flow. The crucial observation is that reversing the flow is equivalent to reflection in a vertical plane, denoted as $x \leftrightarrow -x$. If the particles would be perfectly aligned with the flow, they would respect this $x \leftrightarrow -x$ symmetry, and the system would be symmetric with respect to a sudden reversal of the primary flow: hence no inversion of the secondary flow could then take place (Fig.~\ref{rice_refl}(a)). Therefore, the strong effect of reversal shows that, in steady state, the packing must be unequal to its mirror image: {\em misalignment} of the particles and flow must be key.

Second, we consider the steady state flow. As the steady state secondary flow is always upwards in the center of the cell,
irrespective of the shear direction, the particles orientation must be evolving after the flow is reversed. Hence, the scenario is as follows: flow leads to orientation of the particles to a reflection-symmetric broken state.
Immediately after reversal, the orientation and ordering of the particles has not changed yet, and the secondary flow reverses. After some time has passed, the particles orientations
adapt to the new primary flow direction, and the secondary flow becomes upward in the center again.

To understand in more detail how misalignment, flow reversal, and symmetries are related, let us consider two types of misalignment --- in the horizontal plane (Fig.8(b)), and out-of-plane (Fig.8(c)). Let us for now assume that the system also possesses
an (approximate) $z\leftrightarrow -z$ symmetry. This symmetry reverses the secondary flow, but leaves the horizontal misalignment unaffected (Fig.8(b)) --- hence, in the presence of $z$ reflection symmetry, the secondary flow must be absent. Of course, the $z$-symmetry of the system is (weakly) broken by both gravity and the shape of the shearing zone. We cannot rule out that this causes the secondary flow, but note
that the flow profiles in split bottom cells are rather unaffected by the direction of gravity, as could be expected for slow flows \cite{ries}.

Let us now consider out-of-plane misalignment of the particles. In contrast to the case of in-plane misalignment, there is no symmetry that prohibits
the secondary flow. Moreover, by a combination of
flow reversal, equivalent to $x \leftrightarrow -x$, and vertical reflection $z \rightarrow -z$, Fig.8(c) illustrates that this keeps the particles out-of-plane orientation, but reverse both the direction of the secondary and primary flow  --- exactly as what happens during shear reversal. This strongly suggests that out of plane misalignment is necessary for driving the secondary flow.

\begin{figure*}[t]
\begin{center}
\includegraphics[width=0.9\textwidth]{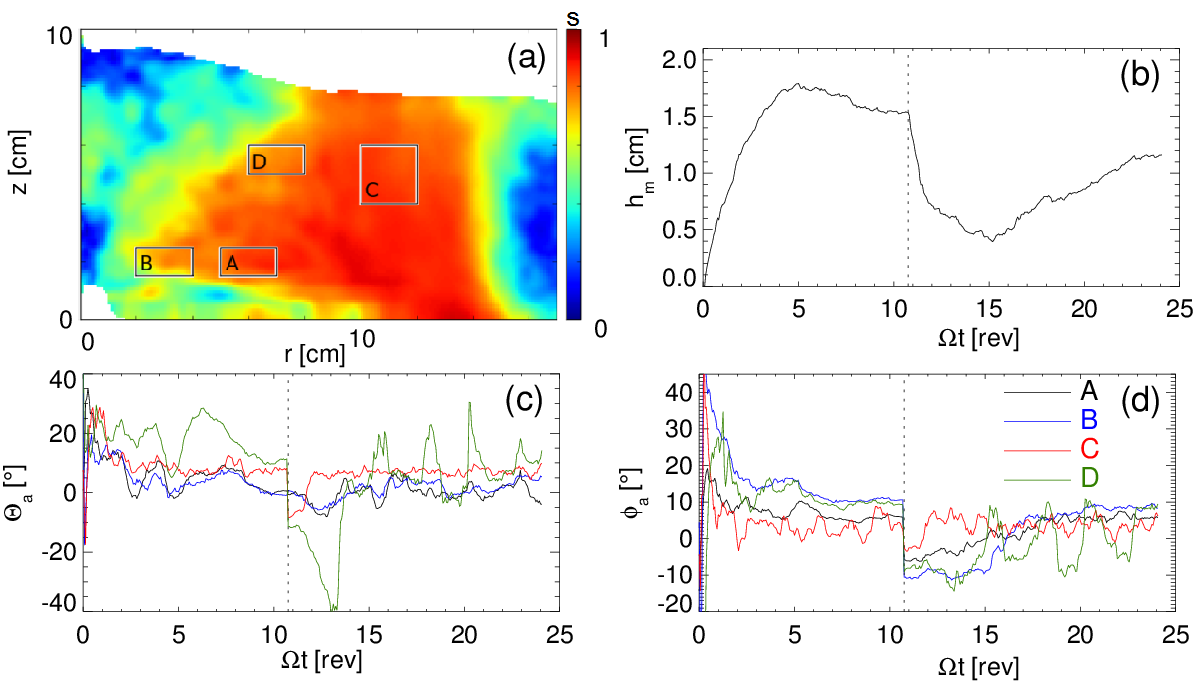}
\caption{{\footnotesize (a) Indication of regions A,B,C and D. The color represents {the orientational order parameter $S$, averaged between 5 and 10.4 revolutions of the plate}. (b) The heap height $h_m$ as a function of disk revolutions for a reversal run. (c) The average horizontal deviation angle $\Theta_a$ in the four regions as indicated in (a). (d) The average vertical deviation angle $\phi_a$. The black dashed line indicates the moment when the flow direction is reversed. It can be seen that different parts of the system take a different time to reorient. However, the reorientation times, in particular of $\phi_a$, correspond well to the time it takes for the heap to start to regrow. }}\label{r_deviation}
\end{center}
\end{figure*}

To probe the particles alignment, we determine the {orientational order tensor $T$ from our 3D data \cite{tamasPRL12,tamasPRE12,balazsPRE14}, and quantify the orientational order parameter $S$, defined as the largest eigenvalue of $T$}. In Fig.~\ref{r_deviation} {we show that $S$, which is a measure for the strength of the orientational order} of the particles, is {largest} in the shear zone --- flow is necessary to align the particles. By combining our {particle tracking} data and $T$ we can also determine the mean misalignment angles between the particles and flow: the in-plane angle
$\Theta_{a}$ {(Fig.8(b))} and out-of-plane angle $\phi_a$ {(Fig.8(c))}. As the
misalignment angles vary throughout the cell, we have probed these in four regions as indicated in Fig.~\ref{r_deviation}(a) during a flow reversal experiment --- note that this is data which is averaged over the azimuthal coordinates in the X-ray tomogram.

In Fig.~\ref{r_deviation}b, we show the heap height $h_m$ as function of {the number of revolutions}, where the dashed line indicates shear reversal --- consistent with earlier data, we see that after reversal,  the heap disappears and reforms over a few {revolutions} of the disk. We note that in comparison to the experiments with rice grains of similar aspect ratio, here both the decay of the heap and its reformation are noticeably slower.
In Fig.~\ref{r_deviation}(c-d) we plot $\Theta_{a}$ and $\phi_a$ for the four regions A-D. By definition, the angles change sign when the flow direction is reversed. We are now interested in the evolution of the mis-match angles. What is striking is that the in-plane mismatch angles $\theta_a$ do not seem to exhibit a systematic trend, whereas the out-of-place angles $\phi_a$ in A, B and D exhibit significant evolution of a strain scale compatible to the strain scale needed to regrow the heap. This suggests strongly that these out-of-plane components evolve together with the magnitude of the secondary flow, as suggested by our symmetry arguments. Note that in area C, in the middle of the shear zone, the misalignment is weaker and without such a clear strain scale. This suggest that the edges of the flowing zone (A, B, D) are most important for setting the secondary flow.

\section{Discussion and Conclusion}
We note that the heaping effect observed here is reminiscent of the so-called \emph{Weissenberg} or rod-climbing effect, which can be observed when
a spinning rod is inserted into a polymer solution:  the fluid will climb up the rod, due to normal stress effects~\cite{Weissenberg}. Surprisingly, viscoelastic fluids exhibit the formation of a heap when driven in a split-bottom geometry~\cite{soto}. However, in both cases, the magnitude of the surface deformation depends on the driving rate and disappears for slow flows --- in contrast to the heaping observed here.

In conclusion, in this paper we have shown that when sufficiently deep layers of granular rods are sheared in a split-bottom geometry, a heap arises at the surface of the packing. The heaping is strongest when axial shear is present near the surface, and we present
strong evidence that heaping is caused by a secondary convective flow. We have presented a symmetry breaking argument, and experimental data that strongly suggest that the convection is the result of an out-of-plane misalignment between the mean orientation of the particles and the streamlines of the flow. This argument also correctly captures the transient disappearance of the heap upon reversal. Our work points to surprisingly strong collective flow phenomena in ensembles of elongated grains, that are missing in current descriptions of granular flows \cite{kamrin}.

\section{Acknowledgments}
We acknowledge technical assistance of Jeroen Mesman, Arthur Blommers for performing the experiments reported in Fig.~\ref{removal} and Fig.~\ref{reversal}, and Elie Wandersman for helpful discussion. GW acknowledges funding by the Dutch funding agency FOM.
{We thank G. Rose and the Inka Lab of Otto von Guericke University Magdeburg for the opportunity to use the X-ray CT facilities. Financial support by the DAAD/M\"OB researcher exchange program (grant no. 29480), the Hungarian Scientific Research Fund (grant no. OTKA NN 107737), and the J\'anos Bolyai Research Scholarship of the Hungarian Academy of Sciences is acknowledged.}

\end{document}